This is a preprint of an article accepted July 9, 2008 for publication in the *Journal of the American Society for Information Science and Technology* 59(13): 2186-2188, copyright 2008.
DOI:http://dx.doi.org/10.1002/asi.20943

# Eigenfactor : Does the Principle of Repeated Improvement Result in Better Journal Impact Estimates than Raw Citation Counts?

Philip M. Davis

Department of Communication

336 Kennedy Hall

Cornell University, Ithaca, NY 14853

email: pmd8@cornell.edu

phone: 607 255-2124

**Abstract**

Eigenfactor.org, a journal evaluation tool which uses an iterative algorithm to weight citations (similar to the PageRank algorithm used for Google) has been proposed as a more valid method for calculating the impact of journals. The purpose of this brief communication is to investigate whether the principle of repeated improvement provides different rankings of journals than does a simple unweighted citation count (the method used by ISI).

**Introduction**

Through the process of referencing other people's ideas, citations create a massive network of scientific papers (Price, 1965). By analyzing this network of citations, one can better understand the origins and history of ideas as they disseminate through the scientific community (Garfield, 1955). In scientific publications, citations perform two distinct functions: they provide a link to a previously published document, and secondly, they perform an acknowledgement of intellectual indebtedness (or credit) to the cited author. It is this second function that is key to the reward system of academic publishing. Sociologist Robert K. Merton referred to citations as a "pellets of peer recognition that aggregate into reputational wealth" (Merton, 1988). Like democratic elections, a citation is like a vote, and those articles, journals or individuals who amass more votes are considered more prestigious.

Yet, measuring citations as 'votes' assumes that each citation has equal worth. A citation from an article published in the journal *Nature* is worth the same as a citation from an article published in an obscure journal. In reality, some citations are clearly more valuable than others (Cronin, 1984).

While the idea of weighting the influence of some journals more than others is not new, Pinski and Narin (1976) are credited with developing the iterative algorithm of calculating influence weights for citing articles based on the number of times that they have been cited. Brin and Page (1998) apply the notion of weighted citations in their PageRank algorithm to calculate the importance of web pages.

Recently, a similar iterative weighting approach has been used by the website, Eigenfactor.org, to calculate the impact of scholarly journals (Bergstrom, 2007a, 2007b). Using data primarily from ISI, the source of the Science Citation Index (SCI) and Journal Citation Reports (JCR), Eigenfactor.org calculates an importance variable (called an Eigenfactor) for each journal. The purpose of this brief communication is to investigate whether the iterative weighting process provides different rankings of journals than using simple unweighted citation counts (the method used by ISI).

**Methods**

The set of 171 journals from the category Medicine (General and Internal) were downloaded from the 2006 edition of JCR. Corresponding Eigenfactors were looked up from Eigenfactor.org. Six journals were removed because they did not have an Eigenfactor, leaving a set of 165 journals. These 165 journals are plotted against total citation counts and Impact Factors from ISI (Figure 1). ISI calculates Journal Impact Factors by dividing the total number of citations a journal receives over the last two years by the total number of articles published in those two years. Essentially, it provides an indicator of citation impact normalized by the size of the journal.

**Results**

Figure 1 shows that total unweighted citation counts are highly correlated with journal Eigenfactors (Pearson rho=0.95). Impact Factors are less closely correlated with Eigenfactors (Pearson rho=0.86), although the relationship is still very strong. It should be noted that the ISI Impact Factor uses a 2-year citation average while the Eigenfactor uses a 5-year average. Generally, longer averages result in less variability.

**Figure 1. Eigenfactors plotted against raw citation counts and Impact Factors for 165 medical journals**

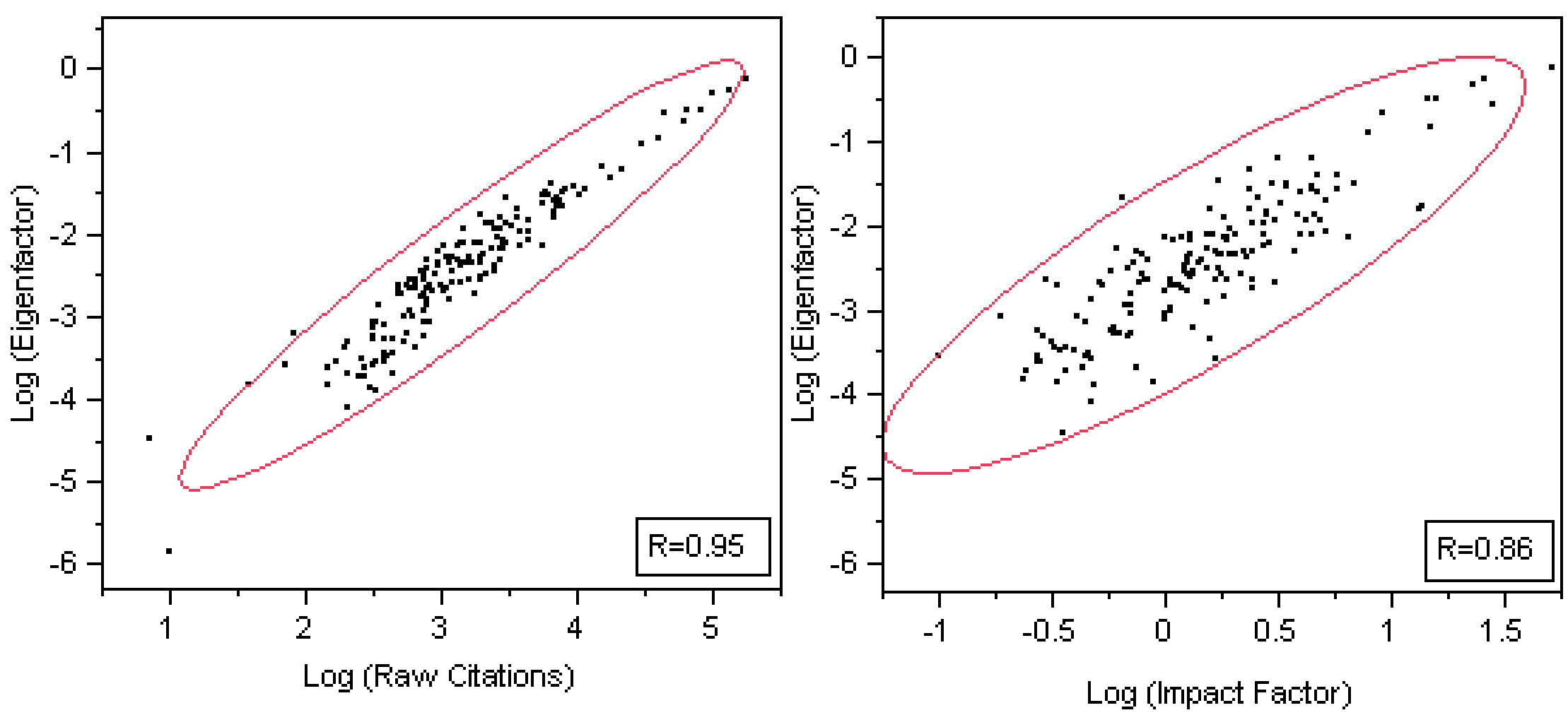

Notes for Figure 1:

1. Data are plotted with 95% bivariate normal density ellipses

2. The logarithm is taken for each variable to conform to the assumptions for Pearson correlation, namely: normality, linearity and homoscedasticity.

In addition, the ordering of journals does not change drastically when journals are ranked by Eigenfactor rank, Total Citation rank, and to a lesser degree, Impact Factor rank. The rank order correlation between Eigenfactor and Total Citations remains very strong (Spearman rho =0.95), and strong between Eigenfactor and Impact Factor (Spearman rho=0.84). The top 20 medical journals are presented in Table 1. As we move down the journal list, the difference in Eigenfactor between consecutively ranked journals becomes smaller and smaller. We should not assume that differences in the third or fourth digit represent meaningful statistical differences.

**Table 1. Top 20 journals in Medicine ranked by Eigenfactor, total citations, and Impact Factor**

| Journal | Eigenfactor | Citations | Impact Factor | Eigen Rank | Citation Rank | Impact Factor Rank |
|---|---|---|---|---|---|---|
| NEW ENGL J MED | 0.7183 | 177505 | 51.296 | 1 | 1 | 1 |
| LANCET | 0.5002 | 133932 | 25.8 | 2 | 2 | 3 |
| JAMA-J AM MED ASSOC | 0.45493 | 100317 | 23.175 | 3 | 3 | 4 |
| J EXP MED | 0.29811 | 65399 | 14.484 | 4 | 5 | 7 |
| J CLIN INVEST | 0.29164 | 80963 | 15.754 | 5 | 4 | 5 |
| NAT MED | 0.26509 | 43664 | 28.588 | 6 | 7 | 2 |
| BRIT MED J | 0.20597 | 61517 | 9.245 | 7 | 6 | 10 |
| ANN INTERN MED | 0.13643 | 39609 | 14.78 | 8 | 8 | 6 |
| ARCH INTERN MED | 0.11489 | 29480 | 7.92 | 9 | 9 | 11 |
| VACCINE | 0.059779 | 15193 | 3.159 | 10 | 12 | 34 |
| AM J MED | 0.056634 | 21290 | 4.518 | 11 | 10 | 22 |
| LIFE SCI | 0.04394 | 17807 | 2.389 | 12 | 11 | 50 |
| MOL THER | 0.037866 | 6397 | 5.841 | 13 | 25 | 15 |
| GENE THER | 0.035742 | 9350 | 4.782 | 14 | 15 | 19 |
| LARYNGOSCOPE | 0.031601 | 11341 | 1.736 | 15 | 13 | 68 |
| STAT MED | 0.030887 | 8376 | 1.737 | 16 | 16 | 67 |
| AM J PREV MED | 0.028953 | 5764 | 3.497 | 17 | 27 | 31 |
| CAN MED ASSOC J | 0.028916 | 7724 | 6.862 | 18 | 17 | 12 |
| J GEN INTERN MED | 0.028292 | 6066 | 2.964 | 19 | 26 | 37 |
| LAB INVEST | 0.027358 | 10307 | 4.453 | 20 | 14 | 25 |

The distribution of citations to scientific journals is extremely skewed (Seglen, 1992). A small number of journals garner the vast majority of citations. As illustrated in Table 1, the *New England Journal of Medicine* alone (out of 165 journals in its category) received 16% of all citations. The top 5 journals (while representing only 3% of the journals) contributed over half (51%) of all the citations.

At least for medical journals, it does not appear that iterative weighting of journals based on citation counts results in rankings that are significantly different from raw citation counts. Or stated another way, the concepts of *popularity* (as measured by total citation counts) and *prestige* (as measured by a weighting mechanism) appear to provide very similar information.